\documentstyle[aps,epsfig]{revtex}
\begin{document}

\title{Dualization of Interacting Theories \\
Including $p=d-1$ Limiting Cases\\ }

\author{S. Ansoldi\footnote{e-mail address:ansoldi@trieste.infn.it}}
\address{Dipartimento di Fisica Teorica\break
Universit\`a di Trieste,\\
INFN, Sezione di Trieste}
\author{A. Aurilia\footnote{e-mail address:aaurilia@csupomona.edu}}
\address{Department of Physics,
California State Polytechnic University\break Pomona, CA 91768}
\author{A. Smailagic \footnote{e-mail address: anais@etfos.hr} }
\address{Department of Physics, University of Osijek, Croatia }
\author{E. Spallucci\footnote{e-mail
address:spallucci@trieste.infn.it}}
\address{Dipartimento di Fisica Teorica\break
Universit\`a di Trieste,\\
INFN, Sezione di Trieste}

\maketitle

\begin{abstract}
We study the vacuum partition functional $Z\left[\, J\,\right]$
for a system of closed, bosonic $p$--branes
coupled to $p$--forms in the limiting case : $p+1=$ spacetime
dimension. We suggest an extension of the duality transformation
which can be applied to the limiting case even though no dual gauge
potential exists in the conventional sense.
The dual action thus obtained describes a current--current, static interaction
within the bulk volume bounded by the $d-1$--brane. Guided by these
results, we then construct a general expression for the parent
Lagrangian that allows for a unified treatment of $p$--duality, even in the
presence of external currents, using a first order formalism instead of the
Bianchi identities. Finally, we show how this generalized dualization approach
can accommodate the inclusion of a massive topological term in the parent
action of an Abelian gauge theory.
\end{abstract}
\vspace*{1cm}
\rightline{\small \bf Accepted for publication in Phys.Lett.B on 99-11-05}
\newcommand\beq{\begin{equation}}
\newcommand\eeq{\end{equation}}
\newcommand\beqa{\begin{eqnarray}}
\newcommand\eeqa{\end{eqnarray}}
\newpage
\section{Introduction}
Recently, within a newly proposed spacetime approach to dualities
\cite{gris}, an interesting new result emerged for the dual theory of
an Abelian gauge field in two dimensions\cite{ae}. The novelty of that
two dimensional model stems from the fact that there is no known dual
version of it within the conventional $p$--duality approach \cite{ct}.
The root of the problem is that a gauge field in two
dimensions corresponds to the limiting value  $p=d-1$, and the usual
dualization procedure is not
applicable because the corresponding Bianchi identities are undefined.
\footnote{ Here $p$ stands for the number of world indices, or rank,
of the gauge potential, and $d$ represents the number of spacetime
dimensions. The   metric is Minkowskian and our signature convention is  
$(\, -\,+\,+\,+\,\dots\, +\,)$.}
As a matter of fact, according to the conventional method, the dual
theory of a $p$--form  involves a $d-p-2$ field, and this correspondence leads 
to the constraint $p\le d-2$.
Nevertheless, one wonders if  there is a way to extend the
$p$--duality approach to the case $p=d-1$, even though the dual field theory 
of the starting matter field is generally undefined. To be sure, in the absence
of interactions, the dual theory of a gauge field strength of maximum
rank $d$ corresponds to a  background field which is constant over the
spacetime manifold. In this narrow sense, one may speak of ``constant--tensor
duality'' for the free theory in the limiting case. However, in a flat
spacetime, such a constant can be gauged away on account of
translational invariance, so that the resulting free theory is essentially
empty\footnote{In a Riemannian spacetime, with non zero curvature,
that arbitrary constant cannot be set, in general, equal to zero, and plays
the physical role of a ``cosmological constant\cite{acl}. As a matter of
fact, the role of that constant is of paramount importance in most
models of cosmic inflation''\cite{akms}.}. It
seems clear, therefore, that the extension of the $p$--duality
approach must go beyond the case of a free theory in the limiting case.
Why would these limiting cases be of any interest? Apart from the
recognized importance of dualities in connection with the theory of
extended objects\cite{lind}, it turns out that such limiting
theories have been shown to be of some phenomenological relevance in
relation to the problem  of confinement\cite{aurilia}, \cite{lush} and
glueball formation \cite{russ}. The question arises, then, as to what happens 
in the case of an interacting theory, for instance in the simplest case in
which a coupling to an {\it external}  current is present.\\
In order to address the above question, let us consider the gauge
invariant action  for a  gauge  field $A$, of rank $p=d-1$ interacting with 
an external current $j$ in a $d$--dimensional Minkowski spacetime,
\beq
S\equiv -{1\over  2d! }\int d^d x \, F_{\,\mu_1\dots\mu_d \,}
\,  F^{\,\mu_1\dots\mu_d \,} -\frac{e}{(d-1)!}
\int d^{\, d}x\,A_{\,\mu_1\dots \mu_{d-1}\,}(x)\,
 j^{\,\mu_1\dots \mu_{d-1}\,}\left(\, x\, ;\partial V \, \right) \ .
\label{uno}
\eeq
The external current
\begin{equation}
j^{\,\mu_1\dots\mu_{d-1}\,}\left(\, x\, ; \partial V \,\right)
\equiv\int_{\partial V} \delta^{d)}\left[\, x - y(\sigma)\,
\right] \, dy^{\mu_1}\wedge \dots\wedge dy^{\mu_{d-1}}
\end{equation}
may be thought of as originating from the timelike history of a
$d-2$--brane, and can be viewed as the boundary current of a
$d$--dimensional {\it bulk} volume, or ``~bag~''.
Accordingly, the boundary $\partial V $ is parametrized by $d-1$
coordinates $\{\,\sigma^a\, , a=1, \dots\, d-1\, \}$.\\
 Gauge invariance of the action requires the current  to satisfy the condition
 \beq
\partial_{\mu_1}j^{\,\mu_1\dots\mu_{d-1}\,}=0 \ .
\label{cons}
\eeq
 The conservation of the boundary current, in turn, implies that $
j^{\,\mu_1\dots\mu_{d-1}\,} $ can be written
 as the divergence of the {\it bulk current} $J$ :
\beq
j^{\,\mu_1\dots\mu_{d-1}\,}\left(\, x\, ; \partial V \,\right)\equiv
\partial_{\mu_d}J^{\mu_1\,\mu_2\dots\mu_d}\left(\, x\, ;  V \,\right)
\label{divj}
\eeq
where
\beq
J^{\mu_1\,\mu_2\dots\mu_d}\left(\, x\, ;  V \,\right)\equiv
\int_V d^d\xi \,\delta^d\left[\, x-z(\xi)\,\right]
dz^{\mu_1}\wedge\dots \wedge dz^{\mu_d} \ .
\label{bagcurr}
\eeq
In the case of the extremal theory encoded in the action functional
(\ref{uno}), the field strength $F$ is an antisymmetric tensor of
maximum rank $d$ for which no Bianchi Identities can be formulated. We propose
to circumvent  this obstacle by using  the {\it first order} formalism. 
Technically, this means that the dualization procedure should be
implemented using the constraint $\delta\left[\, F - dA\, \right]$ instead of
$\delta\left[\, dF \,\right]$ within the path integral approach. In
other words, as in the case of electrodynamics in two spacetime dimensions,
we impose that the ``~Maxwell tensor~'' is, in fact, the covariant curl of
the gauge potential. This leads us to the {\it partition functional}
corresponding to the original action (\ref{uno}),
\beqa
&& Z[j]=  Z[0]^{-1}\int
[dF][dA]\,\delta\left[\, F_{\mu_1\dots\mu_d }-\partial_{[\,\mu_1}
A_{\mu_2\dots \mu_d\, ]}\, \right]\times\nonumber\\
&&\times\exp\left\{{i\over  2  \, d\, !  }\int d^4 x \,
 F^{\mu_1\dots\mu_d }\,F_{\mu_1\dots\mu_d  }
-{ie\over (d-1)!}\int d^dx \, A_{\,\mu_1\dots\mu_{d-1}\,}(x)\,
j^{\,\mu_1\dots\mu_{d-1}\,}\left(\, x\, ;\partial V \right)\right\}  \ .
\label{bag}
\eeqa
Next, we write the Dirac--delta function  in the exponential
form with the help of a Lagrange multiplier $ B^{\,\mu_1\dots\mu_d\,}\,(x)$,
\beq
\delta\left[\, F_{\mu_1\dots\mu_d}-
\partial_{[\,\mu_1}\, A_{\mu_2\dots\mu_d\,]}\, \right]=
\int [dB]\exp\left\{-{i\over d\, !}\int d^d x\,
\, B^{\mu_1\dots\mu_d}\,\left(\, F_{\, \mu_1\dots\mu_d\, }-
\partial_{[\,\mu_1 }\, A_{\mu_2 \dots\mu_d\,]}\,\right)\, \right\}\ .
\eeq
>From here, the {\it dual}  partition functional is obtained by
integrating out independently both $F$ and $A$.\\
First, the path integral (\ref{bag}), being Gaussian in  $F$, is
easily evaluated, and one finds
\beqa
&&\int [\, dF\, ] \exp\left\{{i\over 2 d\, !}\int d^4 x \,\left[\,
\, F^{ \, \mu_1\dots\mu_d\, }\, F_{\, \mu_1\dots\mu_d\, }+
2F^{ \, \mu_1\dots\mu_d\, }\, \,
B_{\,\mu_1\dots\mu_d\,}\, \right]\, \right\}=\nonumber\\
&&=\exp\left\{\, {i\over 2d!}\int d^4x\,
B_{\,\mu_1\dots\mu_d\,}\,B^{\,\mu_1\dots\mu_d\,}\, \right\}  .
\eeqa
As for the $A$--integration, we note that the key feature of the first
order formalism is to introduce  the original gauge potential $A$ {\it linearly}
into the path integral. In this way, the potential $A$ appears as an additional
Lagrange multiplier which, after integration, yields the following condition

\beqa
&&\int [dA]\exp\left\{\,{i\over (d-1)!}
\int d^4x \,\, A_{\mu_1\dots \mu_{d-1}}\,\left(\,
\partial_{\mu_1} \, B^{\mu_1\mu_2\dots \mu_d } - \,e\,
j^{\mu_2\dots \mu_d }\, \right)\,\right\}= \nonumber\\
&&=\delta\left[\,\partial_{\mu_1}  \, B^{\mu_1\mu_2\dots \mu_d }
-\,e\,  j^{\mu_2\dots \mu_d } \,\right] \ .
\label{bo}
\eeqa
The effect of the above delta--function is to restrict the
``~trajectories~''in the path integral to the family of {\it classical field 
equation} for the dual field $B$. In order to extract the physical meaning of 
the dual theory, it may be helpful to rewrite the above delta function in terms 
of the bulk current $J$,  as follows
\beq
\delta\left[\,\partial_{\mu_1 } B^{ \mu_1\mu_2\dots \mu_d} -e\,
j^{\mu_2\dots \mu_d }\,\right]=
\left(\,det\,\Box\,\right)^{-1/2}\delta\left[\, B^{\mu_1\mu_2\dots
\mu_d }-B^{\mu_1\mu_2\dots \mu_d }_0
-\,e\, J^{\mu_1\mu_2\dots \mu_d}\,\right]\label{deltak}\  .
\eeq
It should be noted that in trading the boundary current for the bulk current in
the delta function (\ref{deltak}), we have introduced an arbitrary
constant field $B^{\mu_1\mu_2\dots \mu_d }_0$. As mentioned in the introduction,
this arbitrary constant represents the solution of the homogeneous equation
for the $B$--field, and corresponds to a cosmological term in the action.\\
Performing  the integration over $B$ with the help of the above
delta--function leads to the following expression of the dual
partition functional,
\beq
\exp \left\{\,i\, W\left[\,J\,\right]\, \right\}=
\exp\left\{\, {\, i\over 2\cdot d\, !}\int d^dx\,\left(\,
B^{\mu_1\mu_2\dots \mu_d}_0 -e\, J^{\mu_1\dots\mu_d }\, \right)^2\,\right\} \ .
\label{kk}
\eeq
Without loss of generality, our discussion can be simplified by
resetting the constant $B_0$ to zero, and by choosing the normalization factor
$Z[0]$ in such a way to cancel the determinant factor appearing in 
(\ref{deltak}). 
With such redefinitions, equation (\ref{kk}) represents a direct 
current--current interaction {\it within the bulk} which is  dual to
the original theory (\ref{uno}) whose interaction takes place {\it
between elements of the boundary} through the mediating agency of a
$(d-1)$--index potential. \\
>From the above result, one might be deceived into thinking that there
are physical quanta being exchanged between $p$--brane elements in this
limiting case. However, the truth of the matter is that a
$(d-1)$--index potential in $d$--dimensions does not represent a genuine
``~radiation~'' field, in the sense that there are no propagating degrees of 
freedom. As we have emphasized earlier, the field strength, in this case, merely
represents a constant background disguised as a gauge field. However,
if there are no physical degrees of freedom in the original theory, the
same must be true for the dual theory, and this begs the question: what is
the nature of the interaction that we have uncovered here?
In this connection, it is instructive to play the game in reverse, and
write the {\it dual} partition functional
$W\left[\, J\,\right]$  in terms of the boundary current $j$. Using
eq.(\ref{bagcurr}), one finds
\begin{equation}
\exp \left\{\, i\, W\left[\, j\,\right]\,\right\}=\exp\left\{\, i\,
{e^2\over (d-1)!}\int d^dx\, j^{\mu_1\dots\mu_{d-1}}\, \frac{1}{\Box}\,
j_{\mu_1\dots\mu_{d-1}}\,\right\}\ .
\label{jj}
\end{equation}
The boundary current $j$, on the other hand, can be expressed through
its Hodge dual
\begin{equation}
j^{\mu_1\dots\mu_{d-1}}=\epsilon^{\mu_1\dots\mu_{d-1}\lambda}\,
j_\lambda
\label{jdual}
\end{equation}
so that eq. (\ref{jj}) can be rewritten as follows
\begin{equation}
\exp \left\{\, i\, W\left[\, j\,\right]\, \right\}=\exp\left\{\, i\,
{e^{\,2}\over 2}\, \int d^dx\,
j^{\mu}\, \frac{1}{\Box}\, j_{\mu}\,\right\} \ .
\label{j}
\end{equation}
The above expression closely resembles the analogous formula for the
interaction between point charges in four dimensions. However, that
analogy is formal. More to the point, the {\it physical content} of
eq.(\ref{j}) is exactly analogous to that of {\it  ``~electrodynamics~'' in two 
dimensions}.
Indeed, the four vector $j_\mu$, in the limiting case of $d-1$--branes,
has no transverse, spatial, components. This is  due to the conservation
property (\ref{cons}) of the original boundary current. In terms of the Hodge
dual  $j_\mu$, that conservation property implies the following relations
\begin{eqnarray}
&&j_\lambda=\partial_\lambda \,\phi\\
&&j_0 = \partial_0\, {\partial^i\, j_i^L \over \nabla^2}
\end{eqnarray}
where $j_i^L$ represents the longitudinal component of the spatial part of 
$j^\mu$. In terms of that longitudinal component one can rewrite eq.(\ref{j}) 
as follows
\begin{equation}
\exp \left\{\, i\, W\left[\, J\,\right]\,\right\}=
\exp\left\{\,  i\, {e^{\,2}\over 2} \,
\int d^dx\, j^i_L\,{1\over\nabla^2 }\, j^i_L\,\right\}
\label{wl}
\end{equation}
and this equation, in turn, can be rewritten in terms of the original brane 
current
\begin{equation}
\exp \left\{\,i\, W\left[\, J\, \right]\,\right\}=\exp\left\{\, i\,
{e^{\,2}\over 2d!} \,\int d^dx\,
j^{0\mu_2\dots\mu_{d-1}}\,{1\over\nabla^2 }\, j_{0\mu_2\dots\mu_{d-1}}
 \,\right\}
\label{wj}
\end{equation}
While the result (\ref{wj})  emphasizes the role of  the ``~zero--component~'' 
of the brane current, thereby violating manifest covariance, it has the
advantage of describing a static, long range,interaction between
surface elements of the boundary. This is clearly reminiscent of the fact that
the original gauge potential $A$ has no propagating degrees of freedom,
much as electromagnetism in {\it two dimensions,} and actually represents its 
generalization for extended objects in higher dimensions. With
hindsight, ``~electrodynamics in two dimensions~'' may be reinterpreted as a
theory of two dimensional bags\cite{aurilia}. \\
Summing up our discussion so far, we have shown how to extend the 
$p$--dualization procedure, within a path integral approach, using a
first order formalism instead of Bianchi Identities in order to include the
limiting case of rank  $p=d-1$ fields. Equation (\ref{kk}) shows that the 
absence of a dual potential results in a {\it local,} (contact) interaction of 
the bulk current $J$ in the dual theory, while the  original potential $A$ 
induces a {\it non--local,} though static, long range interaction on the 
boundary. The success of the procedure relies on the fact that it dualizes the
field strength $F(A)$ to a  field $B$, which, by itself, is not necessarily
the covariant curl of a dual potential. In this sense, even if the  dual
field of the gauge potential $A$ does not exist, one can still construct a dual
theory for its field strength. This would be impossible using the second order
formalism since, in that formalism, it is the gauge potential $A$ that
is dualized. Furthermore, it can be seen from eq.(\ref{wj}) that a
physically meaningful dual theory exist only in case of an interacting theory, 
as anticipated in the introduction.
With the above results in hand, we turn now to the following question
that arises naturally from our preceding discussion: can one include the
usual formulation of $p$--duality within the extended dualization procedure
outlined above for the limiting theory ? Clearly, this would be
desirable in order to have a unified approach to $p$--duality for all values of
 $p$.\\
Our purpose in the remaining part of this letter, then, is to show that
it is indeed possible to reformulate the whole $p$--duality approach
without using the Bianchi identities.
The road to a unified formulation starts from the ``~parent Lagrangian~'' for a 
$p$--form $A$ and an external current $K$ (in the case of an interacting theory)
coupled to a field $B$, which we later identify as the {\it dual}
field. Our strategy, then, is to construct the parent Lagrangian  in such a
way that the dualization procedure is applied to the  field strength rather than
the gauge potential $A$. In this approach, the procedure will turn out to
be equivalent to the first order formalism. What remains to be seen is
that in a non--limiting case the new procedure gives the same result as
the standard approach based on Bianchi identities.\\
We take the parent Lagrangian to be of the form \footnote{In the
following discussion we have suppressed all indices in order to simplify the
notation. Thus, the Hodge dual of a $(p+1)$--form, including the  appropriate 
combinatorial factor, is simply indicated by $ F^{\,*}\equiv {1\over
(p+1)!} \epsilon^{\mu_1\dots\mu_{d-p-1}\dots\mu_d} \,
F_{\mu_{d-p-1}\dots\mu_d}$,
while the product of two $p$--forms becomes: $A\, B\equiv {1\over
p!}\,A_{\mu_1\dots \mu_p}\, B^{\mu_1\dots \mu_p}$.} 

\beq
L_P=-{1\over 2}\left(B - e\, J^* \right)^2 + B\, F^{\,*}(A) +g\, K\, B
\label{parent}
\eeq
where we have  introduced both an ``~electric brane~'' current $J$,  a
``~magnetic brane~'' current $K$, and the corresponding ``~electric~''
and ``~magnetic~'' charges $e$ and $g$. The introduction of two distinct
currents is designed to reproduce within our formalism, among other dualitites,
also the well known Dirac electric/magnetic duality. Furthermore, $ F(A)$ is
assumed, at the outset, to be the curl of the gauge potential $p$--form $A$, 
while the 
dual field $B$ is a $(d-p-1)$--form to be determined in the course of
dualization. Our procedure, which is encapsulated in the diagram of Fig.\ref{f1}
involves two distinct steps which we discuss separately:
\begin{figure}
\centerline{\epsfig{figure=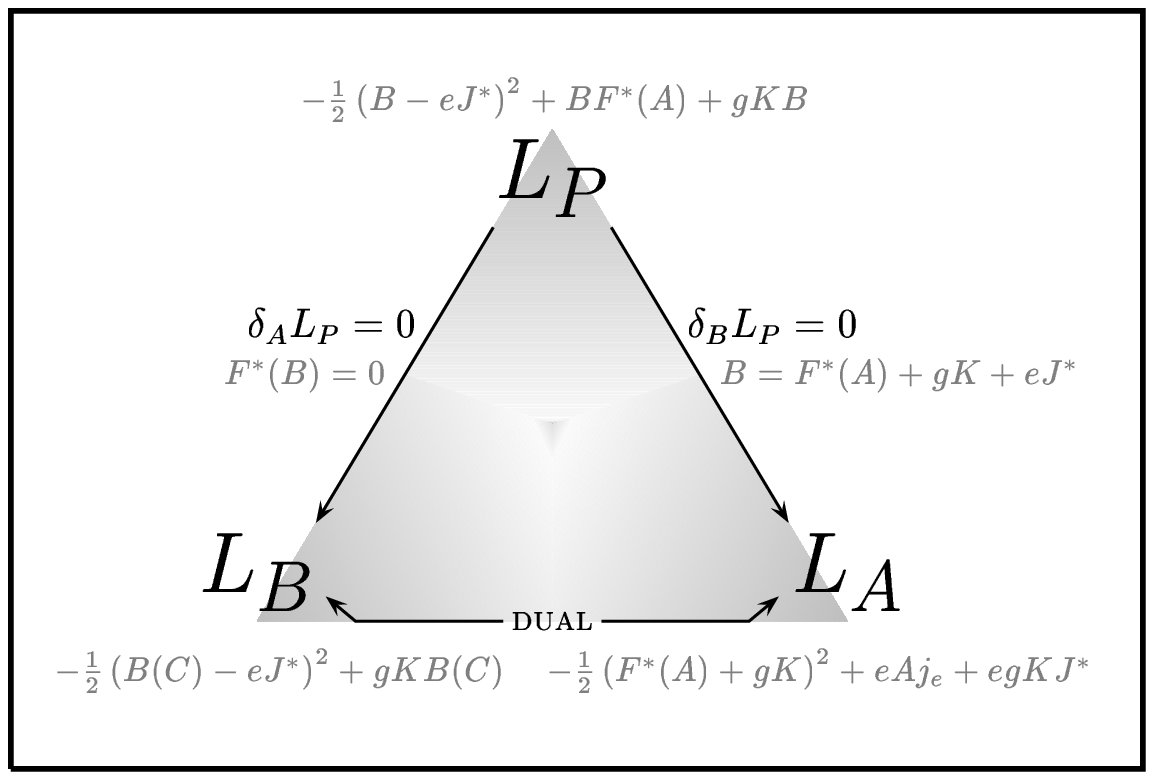,width=11.5cm}}
\caption{}
\label{f1}
\end{figure}
i) variation with respect to the $B$ field in the parent Lagrangian
yields the equation of motion
\beq
\delta_B\, L_P=0\longrightarrow B=  F^{\,*} (A) +g\, K +e\,  J^*
\label{bfk}
\eeq
which, when inserted back into eq.(\ref{parent}), gives the Lagrangian
of the {\it interacting theory} for the field $A$  as follows
\beq
L_A= -{1\over 2}\left(\, F(A) +g(-1)^ { (p+1)(d-p-1) } 
K^* \, \right)^2  +e\, A\, j_e + e\,g\, K \, J^* \ .
\label{la}
\eeq
The ``electric  boundary'' current $j_e $, coupled to the gauge potential
$A$, can be expressed, as before, in terms  of the ``electric'' bulk current $J$
\beq
j_e= \partial J  .
\eeq
Inspection of eq. (\ref{la}) shows that there is a current--current
{\it contact term} $K^2$, as well as a mixed term $K \, J^*$. Such contact
terms are unavoidable consequences of the dualization procedure, and were
noticed for the first time in ref.\cite{dj}.
The  mixed contact term  $K \, J^*$ seems to be especially relevant  
since it leads to the Dirac charge quantization condition
ref.\cite{hk}. Their overall importance for our present discussion will become
clear at the end of this letter.
Next, we redefine the field strength, in order to absorb
the magnetic bulk current: $\widetilde F= F +(-1)^ { (p+1)(d-p-1) } 
 K^{\, *}$ \,\,\,and this
leads to the simple expression for the Lagrangian of the $A$ field
\beq
L_A= -{1\over 2} \widetilde F\, {}^2  +e\, A\, j_e + e\,g\, K J^*
\label{lsing}
\eeq
and to the corresponding field equation
\beq
\partial  \widetilde F= e\, j_e  .
\label{ftilde}
\eeq
ii) The dual theory is obtained by variation of the parent Lagrangian
with respect to $A$,
\beq
\delta_A\, L=0\longrightarrow F^{\,*}(B)= 0   .
\label{bdu}
\eeq
In all cases for which $p\le d-2$, the above ``equation of motion'' ,
it turns out, has a solution which  defines the dual field $B$ as the field
strength of a {\it dual potential $C$}, in agreement with the result of the
standard approach. This proves the equivalence of the two procedures for non
limiting cases. When eq.(\ref{bdu})  is inserted back in the expression
(\ref{parent}), one obtains   the dual theory for the  field $C$ coupled to the
external magnetic current $K$, as described by the Lagrangian
\beq
L_B=-{1\over 2}\left(\, B(C) -e\, J^*\,\right)^2 +g\, K\, B(C)  .
\label{lb}
\eeq
The above Lagrangian leads to the field equations
\beq
\partial\widetilde B = g\,j_m
\label{mag}
\eeq
where, once again, we have used the redefinition $\widetilde B\equiv  B
-e\, J^*$. Similarly, the magnetic boundary current can be expressed as
$j_m=\partial K$.\\
As a check on the consistency of our procedure, it seems worth
observing that equations (\ref{ftilde}) and (\ref{mag}) in the case $p=1$,
$d=4$, reproduce the Dirac electric/magnetic duality. Furthermore, one
can easily check that our procedure reproduces the well known 
{\it scalar--tensor} 
duality between an interacting scalar field in four dimensions, and an
interacting two--index antisymmetric gauge field. Indeed, with the
choice $p=0\quad (\longrightarrow F_\mu(\phi)=\partial_\mu\phi$ $ B(C)
\longrightarrow  B^{\mu\nu\rho}(C)$) and  $J=0$, one finds a Lagrangian
for a scalar field
\beq
L(\phi)= -{1\over 2}\left(\,\partial_\mu\phi + g\, K^*_\mu\,\right)^2
\eeq
which, in turn, leads to the equation of motion
\beq
\Box\,\phi=g\,\partial_\mu K^{*\,\mu}\equiv g\, F^*(K)\label{scal1}
\eeq
while the dual field Lagrangian follows from (\ref{lb})
 \beq
L_C= -{1\over 2\cdot 3!}\, B^{\mu\nu\rho}(C)\, B_{\mu\nu\rho}(C)
+{g\over 3!} \, B_{\mu\nu\rho}(C)\, K^{\mu\nu\rho}
 \eeq
 which gives the equation of motion
 \beq
\partial_\mu B^{\mu\nu\rho}(C)=-g\, j^{\nu\rho}\equiv -g\,
 \partial_\mu K^{\mu\nu\rho}.
 \label{scal2}
 \eeq
Along the same lines one can prove that the above procedure gives all
known dual interacting theories whose current--free version are 
described, for example, in ref.\cite{lind}.
Finally, in order to prove the equivalence to the first order
formalism described previously using the path integral approach, we take the 
limiting value $p=d-1$ and $K=0$ in eq.(\ref{la}). This choice immediately 
leads to

\beq
 L_A=-{1\over 2\cdot d!}\, F^2_{\mu_1\dots\mu_d}(A)
+{e\over (d-1 )!}\, A_{\mu_1\dots\mu_{d-1} } \, j^{\mu_1\dots\mu_{d-1} }  \ .
 \eeq
This is the same expression derived from the action functional
(\ref{uno}).
The dual theory follows instead from eq. (\ref{bdu}). As already mentioned, 
normally that equation defines $B$ as the field strength of some potential $C$,
 except in the limiting case where it gives the condition
\beq
 \epsilon^{\lambda\mu\nu\rho}\, \partial_\rho\, B=0\longrightarrow
 B=const.\equiv B_0  .
 \eeq
This is the same constant field encountered in (\ref{deltak}).
The dual theory is obtained via eq.(\ref{lb}),
 \beq
 L_B=-{1\over 2}\, (B_0- e J^*)^2
 \eeq
which is the same result as in eq. (\ref{kk}) once we identify the Hodge duals
 $B^{\mu_1\dots\mu_d}_0=\epsilon^{\mu_1\dots\mu_d}B_0 $ and $
 J^{\mu_1\dots\mu_d}=\epsilon^{\mu_1\dots\mu_d}J^*$. \\
This shows the asserted equivalence of the path integral and
algebraic procedure in the limiting case, thus providing us with the non
trivial result  that the modified $p$--duality procedure is
applicable to {\it any} interacting theory. As a matter of fact, we
can proceed one step further with our extension of the $p$--duality
procedure, and show that the algebraic procedure includes massive Abelian
topological theories as well. \\
In order to substantiate the above statement, we include in the parent
Lagrangian  a {\it massive topological term,} $m\, A\, F^{\, *}(A)$,
for the $A$ field. The index structure of such topological terms  imposes
the following restriction on the dimensionality of spacetime:
$d=2p+1$, showing, as is well known, that  topological terms can exist
a priori only in odd dimensions. The addition of the topological term in
the parent Lagrangian does not affect equation (\ref{bfk}), which is
obtained by variation of the Lagrangian with respect to the $B$ field. Thus,
inserting eq.(\ref{bfk}) back into the parent Lagrangian yields the following
Lagrangian for the {\it massive topological theory} for the  $A$ field
\beq
 L_A= -{1\over 2}\left(\, F(A) + g\, K^* \, \right)^2 + m\, A\,
F^{\,*}(A)\ .
 \label{latop}
 \eeq
Here we have kept only the ``~magnetic~'' current $K$ for an easier
comparison with the results in ref.(\cite{dj}). In this connection, we also note
that variation of the topological mass term alone with respect to $A$ gives
a term of the form, $m\,[\, 1+(-1)^{p+1}]\, F^{\,*}(A)$. This term contributes 
to
the equation of motion only if $p=2k-1$, while for $p=2k$ the topological
term is a total derivative. Taking into account the previous
restriction to odd
$p$, leads to the number of dimensions $d=4k-1$ quoted in
ref.(\cite{dj}). In a spacetime with such dimensions, the equation of motion 
for $A$ becomes
 \beq
\partial  \widetilde F=  -2m\, F^{\,*}(A) \ .
\label{selfd}
 \eeq
On the other hand, varying the full parent Lagrangian with respect to
$A$, yields  the equation of ``motion'' for the dual field $B$
\beq
 F^{\,*}(B)=-2m\, F^{\,*}(A)
 \label{bad}
 \eeq
which represents an extension of the result (\ref{bdu}) for the topological
term. It is also worth mentioning that eq. (\ref{bdu}) previously was
used to define the dual field as the field strength of the dual potential $C$,
while eq. (\ref{bad}) implies that $B$ has to be of the same rank as
$A$. Inserting eq. (\ref{bad}) into the parent Lagrangian leads to the dual
Lagrangian for the {\it massive ``~topological~'' theory}
\beq
  L_B=-{1\over 2}\, B^2 +g\, K\, B + {1\over 4m}\, B\, F^*(B)
  \label{lbm}
  \eeq
  from which we derive the following equation of motion
\beq
 B= -{1\over 2m}\, F^{\,*} (B) + g\, K  \ .
\label{bselfd}
\eeq
In this way we have shown that the dual version of an {\it interacting,}
topologically massive, Abelian gauge theory discussed, for instance,
in ref.( \cite{djt}), is an integral part of the modified $p$--duality approach, 
thus generalizing the results reported in ref.( \cite{dj}) to arbitrary 
dimensions $d=4k-1$. \\ 
We conclude this letter with a remark on a general property of interacting
dual theories in regards to external currents. As it can be seen from
eq.(\ref{lsing})  the current $j_e$ which is coupled to the gauge
potential $A$ can be expressed in terms of the bulk current as follows,
$j_e=\partial J$. On the other hand, in the
absence of a magnetic current, one can see from (\ref{lb}) that the
dual potential $C$ couples to another {\it electric} current as a
consequence of the dualization procedure. This second current, while implicitly
related to $J$, say $\widetilde j_e=F^*(J)$, is not necessarily given by the 
divergence of the boundary current. Hence, {\it a priori} those two current are 
not related to each other in most theories encompassed by our procedure.
However, an exception to the rule is found in the limiting case $p=d-1$. In such
a case, one can see that the two currents are given by the explicit expressions:
$j_e^ {\mu_2\dots\mu_d} =\epsilon^{\mu_1\mu_2\dots\mu_d}\partial_{\mu_1} J^*$
and $\widetilde j_e^\mu =\partial^\mu\, J^*$, where $J^*$ represents
a zero--form \footnote{We have chosen the symbol $J^*$ for  the zero--form to 
order to match the notation in the parent Lagrangian (\ref{parent}). Note that 
the same reasoning applies to the ``~magnetic~''  current  $K$ in the absence 
of an ``~electric~'' current $J$.}.
This explicit representation of the two currents leads  to the identification 
$\widetilde j_e^\mu= j_e^{*\, \mu}$ which shows that, {\it in the limiting case,
} they are, in fact, related by the operation of Hodge duality.

	\end{document}